\documentclass{sig-alternate}

\usepackage{graphicx}
\usepackage{algorithm}
\usepackage{algorithmic}

\newtheorem{de}{Definition}

\begin{document}

\conferenceinfo{SAC'09}{March 8-12, 2009, Honolulu, Hawaii, U.S.A.}
\CopyrightYear{2009} % Allows default copyright year (2002) to be over-ridden - IF NEED BE.
\crdata{978-1-60558-166-8/09/03}

\title{Enhancing XML Data Warehouse Query Performance\\ by Fragmentation}

\author{Hadj Mahboubi and J\'{e}r\^{o}me Darmont \\
\affaddr{University of Lyon (ERIC)}\\
\affaddr{5 avenue Pierre Mend\`{e}s-France}\\
\affaddr{69676 Bron Cedex} \\ \affaddr{France}\\
\affaddr{\{hadj.mahboubi,jerome.darmont\}@eric.univ-lyon2.fr } }

\maketitle

\begin{abstract}
XML data warehouses form an interesting basis for decision-support
applications that exploit heterogeneous data from multiple sources.
However, XML-native database systems currently suffer from limited
performances in terms of manageable data volume and response time
for complex analytical queries. Fragmenting and distributing XML
data warehouses (e.g., on data grids) allow to address both these
issues. In this paper, we work on XML warehouse fragmentation. In
relational data warehouses, several studies recommend the use of
derived horizontal fragmentation. Hence, we propose to adapt it to
the XML context. We particularly focus on the initial horizontal
fragmentation of dimensions' XML documents and exploit two
alternative algorithms. We experimentally validate our proposal and
compare these alternatives with respect to a unified XML warehouse
model we advocate for.
\end{abstract}


\section{Introduction}
\label{sec:introduction}

Decision-support applications currently exploit more and more
heterogeneous data from various sources. In this context, the
eXtensible Markup Language (XML) is becoming a standard for
representing complex business data \cite{BeyerCCOP05} and can
greatly help in their integration, warehousing and analysis. Many
efforts toward XML data warehousing have indeed been achieved in the
past few years \cite{BoussaidMCA06,Pokorny02}, as well as efforts
for extending the XQuery language with near On-Line Analytical
Processing (OLAP) capabilities such as advanced grouping and
aggregation features \cite{BeyerCCOP05}. This research notably aims
at taking into account specificities of XML data (e.g.,
heterogeneous number and order of dimensions or complex measures in
facts, ragged dimension hierarchies, etc.) that would be intricate
to handle in a relational environment.

XML-native database management systems (DBMSs) supporting XQuery
should form the basic storage component of XML warehouses. However,
they currently present poor performances when dealing with the large
data volumes and complex analytical queries that are typical in data
warehousing. Distributing a warehouse on a grid-like network can
contribute to improve storage and query performance. Such a
framework indeed provides both computing power and distributed
storage resources. Thus, it can be used to handle large data
warehouses efficiently.

Traditionally, the distribution process starts with data
fragmentation. Fragmentation consists in splitting a data set into
two or more parts (fragments) such that the combination of the
fragments yields the original warehouse without any loss nor
addition of information. In the relational context, derived
horizontal fragmentation is acknowledged as best-suited to data
warehouses, because it takes decision-support query requirements
into consideration and avoids computing unnecessary join operations
\cite{BellatrecheB05}. Several approaches have also been proposed
for XML data fragmentation, but they do not take data warehouse
multidimensional architectures (i.e., star-like schemas) into
account.

In this paper, we thus propose to adapt derived horizontal
fragmentation techniques developed for relational data warehouses to
the XML context. We particularly focus on the initial horizontal
fragmentation of dimensions and adapt and compare the two major
algorithms that address this issue: the predicate construction
\cite{NoamanB99} and the affinity-based \cite{NavatheKR95}
strategies.

Adapting these relational techniques onto XML warehouses requires a
well-identified XML warehouse model. Unfortunately, although XML
warehouse architectures from the literature share a lot of concepts
(mostly originating from classical data warehousing), they are
nonetheless all different. Hence, as a secondary contribution of
this paper, we propose a unified, reference    XML data warehouse
model that synthesizes and enhances existing models, and on which we
can base our fragmentation work.

The remainder of this paper is organized as follows. First, we
introduce the state of the art regarding XML data warehouses, as
well as our own reference XML warehouse model
(Section~\ref{sec:XMLDW}). Then, we present general definitions
about fragmentation and discuss existing research related to
relational data warehouse and XML data fragmentation
(Section~\ref{sec:fragmentation}). We detail the specifics of our
adaptation to XML data warehouse fragmentation
(Section~\ref{sec:xmldw_fragmentation}) and experimentally
demonstrate that proper fragmentation significantly reduces the
execution time of analytical XQueries (Section \ref{sec:exps}). We
finally conclude this paper and hint at future research directions
(Section \ref{sec:conclusion}).

\section{XML data warehousing}
\label{sec:XMLDW}

\subsection{Related work} \label{sec:XMLWarehouses}

Several studies address the issue of designing and building XML data
warehouses. They propose to use XML documents to manage or represent
facts and dimensions. The main objective of these approaches is to
enable a native storage of the warehouse and its easy interrogation
with XML query languages.

Pokorn{\'y} models a XML-star schema in XML by defining dimension
hierarchies as sets of logically connected collections of XML data,
and facts as XML data elements \cite{Pokorny02}.
H{\"u}mmer \textit{et al.} propose a family of templates, named
XCube, enabling the description of a multidimensional structure
(dimension and fact data) for integrating several data warehouses
into a virtual or federated warehouse \cite{HummerBH03}.
Rusu \textit{et al.} propose a methodology, based on the XQuery
technology, for building XML data warehouses. This methodology
covers processes such as data cleaning, summarization,
intermediating XML documents, updating/linking existing documents
and creating fact tables \cite{RT05}. Facts and dimensions are
represented by XML documents built with XQueries.
Park \textit{et al.} introduce a framework for the multidimensional
analysis of XML documents, named XML-OLAP \cite{ParkHS05}. XML-OLAP
is based on an XML warehouse where every fact and dimension is
stored as an XML document. The proposed model features a single
repository of XML documents for facts and multiple repositories of
XML documents for dimensions (one repository per dimension).
Eventually, Boussa{\"\i }d \textit{et al.} propose an XML-based
methodology, named X-Warehousing, for warehousing complex data
\cite{BoussaidMCA06}. They use XML Schema as a modeling language to
represent user analysis needs.

\subsection{XML data warehouse reference model} \label{sec:xdwm}

The studies enumerated in Section \ref{sec:XMLWarehouses}, though
all different, more or less converge toward a unified XML warehouse
model. They mostly differ in the way dimensions are handled and the
number of XML documents that are used to store facts and dimensions.
%
%We may distinguish four different families of physical
%architectures: (1) one XML document for storing facts and another
%for storing all dimension-related information (XCube); (2) a
%collection of XML documents that each embed one fact and its related
%dimensions (X-Warehousing); (3) a collection of XML documents where
%facts and dimensions are each stored in one separate document
%(XML-OLAP); (4) one XML document for storing facts and one XML
%document for storing each dimension (analogous to relational
%star-like schemas).
%
A performance evaluation study of these different representations
showed that representing facts in one single XML document and each
dimension in one XML document allowed the best performance
\cite{DoulkiliBB06}. Moreover, this representation allows to model
constellation schemas without duplicating dimension information.
Several fact documents can indeed share the same dimensions.
Furthermore, since each dimension and its hierarchical levels are
stored in one XML document, dimension updates are more easily and
efficiently performed than if dimensions were either embedded with
the facts or all stored in one single document.

Hence, we adopt this architecture model.
%It is actually the
%translation of a classical constellation schema from the relational
%model to the XML model, i.e., tables become XML documents. Note,
%however, that XML warehouses can bear irregular structures that are
%not possible in relational warehouses.
%
More precisely, our reference data warehouse is composed of the
following XML documents (Definition~\ref{def_graph}):
\begin{enumerate}
\item \textit{dw-model.xml} that represents warehouse metadata,
the XML graph representing warehouse metadata is denoted
$G_{dw-model}$;
\item a set of $facts_f.xml$ documents that each
store information related to set of facts $f$;

\item a set of $dimension_{d}.xml$ documents that each store a
given dimension $d$'s member values.
\end{enumerate}

\begin{de}
An XML document is defined as a graph (XML graph) whose nodes
represent document elements or attributes, and edges represent the
element / sub-element (or parent-child) relationship. Edges are
labeled with element or attribute names. \label{def_graph}
\end{de}

\begin{figure}[h]
{\centering
\resizebox*{0.5\textwidth}{!}{\includegraphics{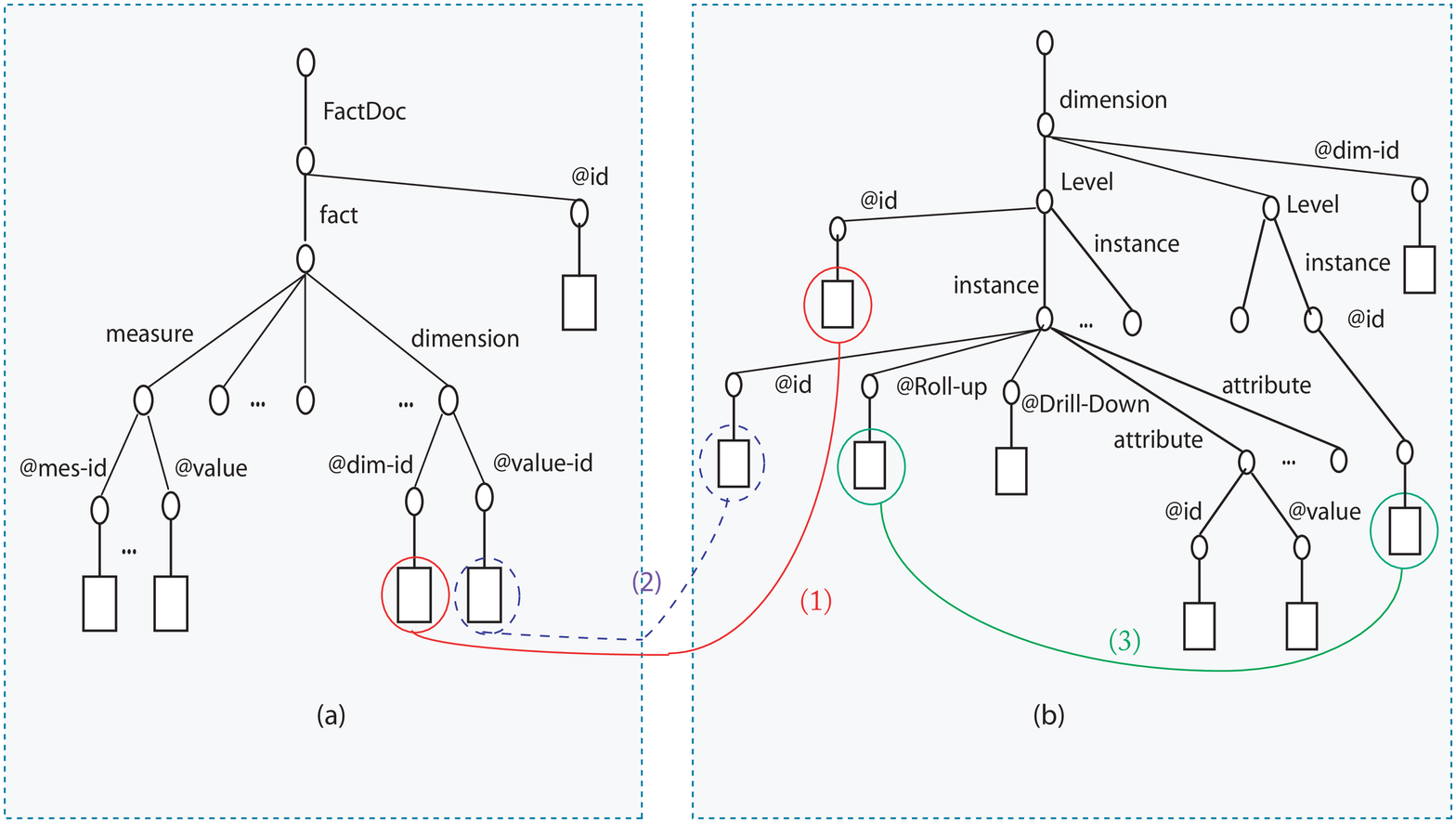}}
\par}
\caption{$facts_f.xml$ (a) and $dimension_{d}.xml$ (b) graph
structure} \label{fig:fact-dimension}
\end{figure}

A $facts_f.xml$ document stores facts
(Figure~\ref{fig:fact-dimension}(a)). The document root node,
\textit{FactDoc}, is composed of \textit{fact} subelements that each
instantiate a fact, i.e., measure values and dimension references.
These identifier-based references support the fact-to-dimension
relationship. The XML graph representing fact set $f$ is denoted
$G_{facts_f}$.

A $dimension_{d}.xml$ document helps instantiate one dimension,
including any hierarchical level
(Figure~\ref{fig:fact-dimension}(b)). Its root node,
\textit{dimension}, is composed of \textit{Level} nodes. Each one
defines a hierarchy level composed of \textit{instance} nodes that
each define the level's member attribute values. In addition, an
\textit{instance} element contains \textit{Roll-up} and
\textit{Drill-Down} attributes that define the hierarchical
relationship within dimension $d$. The XML graph representing
dimension $d$ is denoted $G_{dimension_{d}}$.

\section{Database fragmentation}
\label{sec:fragmentation}

\subsection{Definition}

There are three fragmentation types in the relational context
\cite{BellatrecheB05}: vertical fragmentation, horizontal
fragmentation and hybrid fragmentation.

Vertical fragmentation splits a relation $R$ into sub-relations that
are projections of $R$ with respect to a subset of attributes. It
consists in grouping together attributes that are frequently
accessed by queries. Vertical fragments are built by projection. The
original relation is reconstructed by joining the fragments.

Horizontal fragmentation divides a relation into subsets of tuples
using query predicates. It reduces query processing costs by
minimizing the number of irrelevant accessed instances. Horizontal
fragments are built by selection. The original relation is
reconstructed by fragment union. A variant, derived horizontal
fragmentation, consists in partitioning a relation with respect to
predicates defined on another relation.

Finally, hybrid fragmentation consists of either horizontal
fragments that are subsequently vertically fragmented, or vertical
fragments that are subsequently horizontally fragmented.

\subsection{Data warehouse fragmentation}
\label{sec:relational-fragmentation}

Many research studies address the issue of fragmenting relational
data warehouses either to efficiently process analytical queries or
to distribute the warehouse.

To improve ad-hoc query performance, Datta \textit{et al.} exploit a
vertical fragmentation of facts to build the Cuio index
\cite{DattaRT99}, while Golfarelli \textit{et al.} apply the same
fragmentation on warehouse views \cite{GolfarelliMR99}.
Bellatreche and Boukhalfa apply horizontal fragmentation to a
star-schema \cite{BellatrecheB05}. Their fragmentation strategy is
based on a query workload and exploits a genetic algorithm to select
a optimal partitioning schema that minimizes query cost.
Finally, Wu and Buchmaan recommend to combine horizontal and
vertical fragmentation for query optimization \cite{WuB97}. A fact
table can be horizontally partitioned according to one or more
dimensions, it can also be vertically partitioned according to its
dimension foreign keys.

To distribute a data warehouse, Noaman \textit{et al.} exploit a
top-down strategy that uses horizontal fragmentation
\cite{NoamanB99}. The authors propose an algorithm for deriving
horizontal fragments from the fact table based on queries that are
defined on all dimension tables. Finally, Wehrle \textit{et al.}
propose to distribute and query a warehouse on a computing grid
\cite{WehrleMT05}. They use derived horizontal fragmentation to
split the data warehouse and build a so-called \textit{block of
chunks}, a data set defining a fragment.

In summary, these proposals generally exploit derived horizontal
fragmentation to reduce irrelevant data access rate and efficiently
process join operations across multiple relations
\cite{BellatrecheB05,NoamanB99,WehrleMT05}. In the literature, the
prevalent methods used for derived horizontal fragmentation are the
following \cite{KoreichiB97}.

\begin{itemize}

\item \textbf{Predicate construction}. This method fragments a relation by using a complete and minimal set
of predicates \cite{NoamanB99}. Completeness means that two relation
instances belonging to the same fragment have the same probability
of being accessed by any query. Minimality garantees that there is
no redundancy in predicates.

\item \textbf{Affinity-based fragmentation}.
This method is an adaptation of vertical fragmentation methods to
horizontal fragmentation \cite{NavatheKR95}. It is based on the
predicate affinity concept \cite{ZhangO94}, where affinity defines
query frequency. Specific matrices (predicate usage and affinity
matrices) are exploited to cluster selection predicates. A cluster
is defined as a selection predicate cycle and forms a dimension
graph fragment.

\end{itemize}

\subsection{XML database fragmentation}
\label{sec:xml-fragmentation}

Recently, several fragmentation techniques for XML data have been
proposed. They split an XML document into a new set of XML
documents. Their main objective is either to improve XML query
performance \cite{GertzB03} or to distribute or exchange XML data
over a network \cite{BonifatiMCJ04,BoseF05}.

To fragment XML documents, Ma \textit{et al.} define a new
fragmentation type: \textit{split} \cite{MaS03}, which is inspired
from the oriented-object domain. This fragmentation splits XML
document elements and assigns a reference to each sub-element. The
references are then added to the Document Type Definition (DTD)
defining the XML document.
Andrade \textit{et al.} propose to apply fragmentation to an
homogeneous XML collection \cite{AndradeRBBM06}. They adapt
traditional fragmentation techniques to an XML document collection
and base their proposal on the Tree Logical Class algebra (TLC)
\cite{PaparizosWLJ04}.

%The authors also evaluate these techniques and show that horizontal
%fragmentation provides the best performance.
%Gertz and Bremer also define horizontal and vertical fragmentation
%for an XML document for a distribution purpose. A fragment is
%defined with a path expression language, called \textit{XF}, which
%is derived from XPath. This fragment is obtained by applying an
%\textit{XF} expression on a graph \textit{RG} representing XML data.
Bose and Fegaras use XML fragments for data exchange in a
peer-to-peer network (P2P), called XP2P \cite{BoseF05}. XML
fragments are interrelated and each is uniquely identified by an
\textit{ID}. The authors propose a fragmentation schema, called
\textit{Tag Structure}, to define the structure of data and
fragmentation information. Bonifati \textit{et al.} also define XML
fragments for a P2P framework \cite{BonifatiMCJ04}. An XML fragment
is obtained and identified by a single path expression, a
root-to-node path expression \textit{XP}, and managed on a specific
peer.

In summary, these proposals adapt classical fragmentation methods to
split XML data. An XML fragment is defined and identified by a path
expression \cite{BonifatiMCJ04} or an XML algebra operator
\cite{AndradeRBBM06}. Fragmentation is performed on a single XML
document \cite{MaS03} or on an homogeneous XML collection
\cite{AndradeRBBM06}.

\section{Fragmenting XML data\\ warehouses}
\label{sec:xmldw_fragmentation}

\subsection{Motivation}

Approaches dealing with fragmentation in XML databases adopt only
primary horizontal fragmentation applied onto one XML document
(Section \ref{sec:xml-fragmentation}). They use fragmentation to
minimize XML query expression execution cost. However, in XML data
warehouses, decision-support queries are more complex: they involve
multiple join operations over multiple XML (fact and dimension)
documents. Hence, primary horizontal fragmentation is not adapted in
our context. Relational data warehouse fragmentation approaches
recommend to use derived horizontal fragmentation (Section
\ref{sec:relational-fragmentation}), which is more adapted to
analytical queries. In addition, there are, to the best of our
knowledge, no XML data warehouses fragmentation works in the
literature. In consequence, we propose to adapt horizontal derived
fragmentation to XML data warehouses (Definition \ref{def_dhf}).

\begin{de}
In an XML data warehouse, derived horizontal fragmentation first
splits horizontally $G_{dimension_{d}}$ \\ graphs with respect to a
given workload $W$, and then partitions the $G_{fact_f}$ graphs with
respect to $G_{dimension_{d}}$ fragments. \label{def_dhf}
\end{de}

\subsection{General principle} \label{sec:g_principle}

%Figure \ref{fig:dw-fragmentation} summarizes our derived horizontal
%fragmentation methodology for XML data warehouses.

In our fragmentation methodology, we first apply a primary
horizontal fragmentation onto warehouse dimensions using either the
predicate construction method, denoted PC, or the affinity-based
method, denoted AB (Section \ref{sec:relational-fragmentation}).
Both these methods input selection predicates (Definition
\ref{def_predicate}) from $W$ (Section \ref{sec:selection_pred}). AB
also exploits data access frequencies. Our adaptations of PC and AB
to the XML context are described in Sections \ref{method1} and
\ref{method2}, respectively. Both help fragment $G_{dimension_{d}}$
graphs. Note that we consider both the PC and AB methods to compare
their efficiency, which has never been addressed in the literature
as far as we know. Based on these fragments, we then fragment the
$G_{facts_f}$ graphs and build a fragmentation schema for the whole
XML data warehouse. This process is detailed in Section \ref{fact}.

%\begin{figure}[hbt]
%{\centering
%\resizebox*{0.3\textwidth}{!}{\includegraphics{dw-fragmentation.eps}}
%\par}
%\caption{XML data warehouse fragmentation} \label{fig:dw-fragmentation}
%\end{figure}

\begin{de}
A selection predicate is defined by expression $p := P_{a_k} \theta
[ value \mid \emptyset_{XPath} (P_{a_k}) \mid Q]$, where $P_{a_k}$
and $Q$ are path expressions %(Definition~\ref{def:pathexression})
and $P_{a_k}$ is defined on attribute $a_k$, $\theta \in \{=, <,>,
\leq, \geq, \neq \}$, $value \in D_k$ where $D_k$ is the domain of
$a_k$, and $\emptyset_{XPath}$ is any XPath
function~\footnote{\scriptsize{http://www.w3.org/TR/xpath-functions/}}.
\label{def_predicate}
\end{de}

%\begin{de}
%A path expression $p$ is a sequence $root_{t}/$ $e_{1}/.../
%\{e_{n}/@a_{k}\}$, where $ \{ e_{1},...,e_{n} \} \in E $, and
%$@a_{k}\in A$. Expression $p$ can contain the symbol $*$ that
%indicates any element in $E$ or the symbol $//$ that indicates any
%sequence $e_{i}/.../e_{j}$ where $i<j$. A [i] symbol can be also
%affected to an element $e_{i}$, to indicate its position in the XML
%graph. \label{def:pathexression}
%\end{de}

\subsection{Primary horizontal fragmentation}

\subsubsection{Selection predicate extraction}
\label{sec:selection_pred}

The set $P$ of selection predicates used to fragment the
$G_{dimension_{d}}$ graphs is identified by parsing $W$. For
example, $p_{1}:=\$y/attribute[$@$id='c\_nation\_key']/$
$@value>'15'$ and $p_{2}:= \$y/attribute[$@$id='p\_type']/$
$@value='PROMO BURNISHED COPPER'$ are selection predicates obtained
from query $q_{1}$ in the sample XQuery workload provided in
Figure~\ref{fig:req}.

\begin{figure}[h]
\centering{ \scriptsize{
\begin{tabular}{cl}\hline
$q_{1}$ &   \textbf{for} \$x \textbf{in} //FactDoc/Fact,\\
&           \$y \textbf{in} //dimensions[$@$dim-id='Customer']/Level/instance\\
&           \$z \textbf{in} //dimensions[$@$dim-id='Part']/Level/instance\\
&            \textbf{where} \$y/attribute[$@$id='c\_nation\_key']/@value='13'\\
&            \textbf{and} \$y/attribute[$@$id='p\_type']/@value='PROMO \\
&            BURNISHED COPPER'\\
&          \textbf{and} \$x/dimension[$@$dim-id='Customer']/$@$value-id=\$y/$@$id\\
&          \textbf{and} \$x/dimension[$@$dim-id='Part']/$@$value-id=\$z/$@$id\\
&       \textbf{return} \$x\\ \\

\dots \\

$q_{10}$ &   \textbf{for} \$x \textbf{in} //FactDoc/Fact,\\
&           \$y \textbf{in} //dimensions[$@$dim-id='Customer']/Level/instance\\
&           \$z \textbf{in} //dimensions[$@$dim-id='Date']/Level/instance\\
&            \textbf{where} \$y/attribute[$@$id='c\_nation\_key']/@value$>$'15'\\
&            \textbf{and} \$y/attribute[$@$id='d\_date\_name']/@value='Saturday'\\
&          \textbf{and} \$x/dimension[$@$dim-id='Customer']/$@$value-id=\$y/$@$id\\
&          \textbf{and} \$x/dimension[$@$dim-id='Part']/$@$value-id=\$z/$@$id\\
&       \textbf{return} \$x\\ \\

\hline
\end{tabular}}
} \caption{Workload snapshot}\label{fig:req}
\end{figure}

\subsubsection{PC primary horizontal fragmentation} \label{method1}

\noindent \textbf{Principle}

\noindent Based on selection predicate set $P$ and metadata from
$G_{dw-model}$, PC identifies candidate $G_{dimension_{d}}$ graphs
for fragmentation. A candidate dimension graph $G_{candidate_{d}}$
is a $G_{dimension_{d}}$ graph targeted by workload queries.

For each candidate dimension and its corresponding selection
predicate set $P_{d} \subset P$, a set of complete and minimal
selection predicates $P_{d}^{'}$ is generated with the COM-MIN
algorithm \cite{OzsuV99} that guarantees completeness and minimality
(Section \ref{sec:relational-fragmentation}). PC finally builds from
$P_{d}^{'}$ a set of minterms that horizontally fragment
the $G_{candidate_{d}}$ graphs. \\

\noindent \textbf{Fragmentation methodology}

\begin{enumerate}

\item \textbf{Attribution of selection predicates to dimension XML graphs}.
This step affects to each dimension graph $G_{dimension_{d}}$ its
corresponding selection predicate set $P_{d} \subset P$. $P_{d}$ is
identified from $G_{dw-model}$, which stores for each dimension its
corresponding attributes. Hence, we can identify candidate dimension
graphs ($G_{candidate_{d}}$) for horizontal fragmentation.\\

\textbf{Example}. Predicate $p_2$ contains attribute
$c\_nation\_key$. In $G_{dw-model}$, $c\_nation\_key$ is a member of
the $customer$ dimension. Hence, we identify
$G_{dimension_{customer}}$ as a candidate dimension for
fragmentation.\\

\item \textbf{Selection predicate completeness and minimality}.
\label{step:com-min} In this step, we apply the COM-MIN algorithm,
which inputs $P_{d}$ and outputs a set of complete and minimal
predicates $P^{'}_{d}$. Given $P^{'}_{d}$, the set $M_{d}$ of
minterm predicates is then constructed.\\
$M_{d}=\{m_{i}|m_{i}=\wedge_{q_{j}\in P} q^{*}_{j}\}$, where
$q^{*}_{j}=q_{j}$ or $q^{*}_{j}=\neg q_{j}$, $1\leq j \leq n$,
$1\leq i \leq 2^{n}$ and $n$ represents the number of selection
predicates. A minterm predicate $m_i \in M_{d}$ is the conjunction
of all predicates from $P^{'}_{d}$, taken in natural or negative
form. $m_{1}=p_{1} \wedge \neg p_{2}$ is an example minterm
predicate, where $p_{1}$ and $p_{2}$ are the sample selection
predicates from Section
\ref{sec:g_principle}.\\

\textbf{Example}. Let $P^{'}_{customer}=\{\$y/attribute[@id=$\\
$'c\_nationkey']/@value=13, \$y$ $/attribute[@id=$\\
$'c\_nationkey']/@value>15\}$ be a complete and minimal set obtained
by the COM-MIN algorithm for dimension $customer$. A minterm $m_{1}$
is $\$y/attribute[$ $@id='c\_nationkey']/@value=13$ and $\$y/attribute[$ $@id='c\_nation$ $key']/@value<=15$. \\

\item \textbf{Candidate graph fragmentation}. \label{step:cgfrag}
This step builds primary horizontal fragments from
$G_{candidate_{d}}$. A fragment is obtained by associating to each
minterm predicate $m_{i} \in M_{d}$ the set of nodes in
$G_{candidate_{d}}$ that verifies it.

\textbf{Example}. Minterm $m_1$ is used to fragment\\
$G_{candidate_{customer}}$. %The obtained fragment is showed in Figure \ref{fig:frag}. Each
%\textit{instance} node verifies minterm $m_1$.

%\begin{figure}[hbt]
%{\centering
%\resizebox*{0.5\textwidth}{!}{\includegraphics{dimension-frag.eps}}
%\par}
%\caption{Sample dimension fragment} \label{fig:frag}
%\end{figure}

\end{enumerate}

\subsubsection{AB primary horizontal fragmentation} \label{method2}

\noindent \textbf{Principle}

\noindent AB uses query frequency to build horizontal fragments by
exploiting specific matrices (predicate usage and affinity
matrices). It clusters selection predicates from $P$ by exploiting a
graphical algorithm. A cluster is defined as a selection predicate
cycle and forms a fragment of a $G_{dimension_{d}}$ graph.\\

\noindent \textbf{Fragmentation methodology}

\begin{enumerate}

\item \textbf{Predicate usage matrix construction}.
The predicate usage matrix, $PUM$, is built based on $P$. It defines
predicate usage of each query $q_{i} \in W$. Matrix lines represent
workload queries and columns simple selection predicates from $P$.
General term $PUM(i,j)$ is set to one if $q_{i}$ includes predicate
$p_{j}$ and to zero otherwise. In addition, the usage frequency of
each query $q_{i}$ is stored in a vector $Freq$.

\textbf{Example}. Tables \ref{Matrice} and \ref{Vector} provide
examples of $PUM$ matrix and query frequency vector, respectively.

\begin{table}[htbp]
   \begin{center}
      \begin{tabular}{|c|c|c|c|c|c|c|}

                \hline query/predicate & $p_{1}$ & $p_{2}$ & $p_{3}$ & $p_{4}$ & ... & $p_{n}$\\
                \hline $q_1$   & 1 &   0 & 0   & 0 &    & 0\\
                \hline $q_2$   & 1 &   1 & 0   & 0 &    & 0\\
                \hline ...&  &  &   & & &  \\
                \hline $q_m$   & 1 &   1 & 0   & 0 &    & 1\\
                \hline
                \end{tabular}\\
                \begin{flushleft}
                \begin{footnotesize}
                $n$ represents the number of selection predicates in
                $P$ and $m$ the number of queries in $W$.
                \end{footnotesize}
                \end{flushleft}
      \caption{Sample predicate usage matrix \label{Matrice}}
     \end{center}
\end{table}

\begin{table}[htbp]
   \begin{center}
      \begin{tabular}{|c|c|c|c|}
                \hline $q_{1}$ & $q_{2}$ & ... & $q_{m}$\\
                \hline 10 & 20 & ... & 5\\
                \hline
                \end{tabular}\\
      \caption{Sample query frequency vector \label{Vector}}
     \end{center}
\end{table}

\item \textbf{Predicate affinity matrix construction}
The predicate affinity matrix, $Aff$, is built from the
$PUM$ matrix (Table \ref{Matrice1}). It is a $n\times n$ matrix, where $n$ represents the number of selection predicates in $P$.
$Aff$ matrix cells can contain numeric or nonnumeric ($\Rightarrow$, $\Leftarrow$ and $*$) values.
A numeric value of an $Aff(i,j)$ cell gives the frequency sum of all
queries referencing both predicates $p_{i}$ and $p_{j}$. A value
"$\Rightarrow$" indicates that predicate $p_{i}$ implies predicate
$p_{j}$; a value "$\Leftarrow$" indicates that predicate $p_{j}$
implies predicate $p_{i}$; and a value "$*$" indicates that
predicates $p_{i}$ and $p_{j}$ are similar. Two predicates $p_{i}$
and $p_{j}$ are similar if: (1) they are defined on the same
attribute; (2) there exists a query $q_{i}$ that uses predicates
$p_{i}$ and $p_{c}$ and another query $q_{j}$ which uses predicates
$p_{j}$ and $p_{c}$; and (3) $p_{c}$ is a selection predicate that
is defined on another attribute than the $p_{i}$ and $p_{j}$
predicates \cite{NavatheKR95}.

\textbf{Example}. Table~\ref{Matrice1} shows a example of affinity
matrix.

\begin{table}[htbp]
   \begin{center}
      \begin{tabular}{|c|c|c|c|c|c|c|}
                \hline           & $p_{1}$ & $p_{2}$ &...& $p_{5}$ & ... & $p_{n}$ \\
                \hline $p_{1}$   &   20    &   0     &   & 10 & &   0     \\
                \hline $p_{2}$   &    0    &   30    &   & & &   $\Leftarrow$    \\
                \hline ...       &         &         &   & & &         \\
                \hline $p_{5}$   &   10    &    0    &   & 25 & &    0     \\
                \hline ...       &         &         &   & & &         \\
                \hline $p_{n}$   &    0    &    $\Rightarrow$   &   & 0 & &   5     \\
                \hline
                \end{tabular}\\
      \caption{Sample predicate affinity matrix \label{Matrice1}}
     \end{center}
   \end{table}

\item \textbf{Predicate clustering}.
This step exploits the graphical algorithm proposed by Navathe
\textit{et al.} \cite{NavatheR89} for vertical fragmentation, which
has been been adapted for horizontal fragmentation
\cite{NavatheKR95}. This algorithm inputs $Aff$ and considers it as
a complete graph, $G_{Aff}$. Then, it forms a linearly connected
spanning-tree. A tree node represents a selection predicate $p_{i}$
in $Aff(i,j)$ and an edge $e(p_{i}, p_{j})$ an affinity value. The
algorithm detects and extracts a set of cycles $C$, where a cycle
$c_{i} \in C$ groups selection predicates sharing values in $Aff$.

\textbf{Example}. Figure \ref{fig:example1} shows a sample $G_{Aff}$
that is build from $Aff(i,j)$. $C=\{ c_1, c_2, ..., c_z\}$, where
$z$ represents the number of cycles. $c_1 = \{p_1, p_3, p_5\}$.

\begin{figure}[hbtp]
{\centering
\resizebox*{0.18\textwidth}{!}{\includegraphics{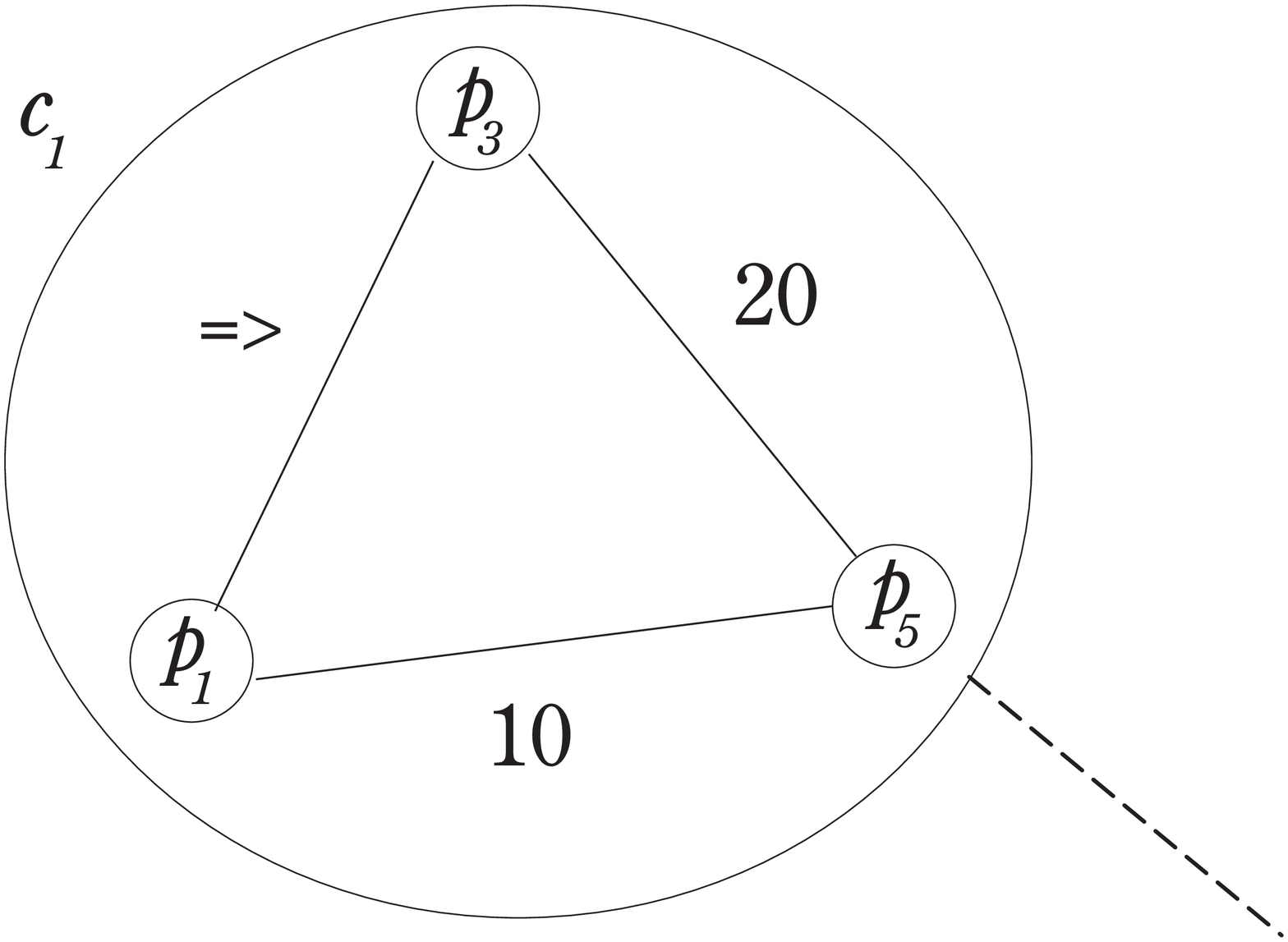}}
\par}
\caption{Predicate clustering example} \label{fig:example1}
\end{figure}

\item \textbf{Compose predicate terms}.
Cycle set $C$ is first evaluated to determine distinct common
attributes in $C$ predicates and construct a specific table called
predicate term schematic table. This table stores attribute usage
for each $c_i$. Based on this table, predicate terms $t_i$ are
constructed. A predicate term $t_i$ constitutes an horizontal entry
in the predicate term schematic table and covers all common
attributes.

\textbf{Example}. Table \ref{table1} gives an example of predicate
term schematic table. Predicates in cycle $c_1$ do not include
attribute $a_{2}$. $c_1$ is hence divided to a set of sub-cycle
$c_{1j}$. Each $c_{1j}$ sub-cycle contains predicates from $c_1$ and
a predicate $p_j$ that includes attribute $a_2$. $t_1 = p_1 \wedge
p_3
\wedge p_5 \wedge p_2$ for $j=2$ is an example of predicate term.\\

\begin{table}[htbp]
   \begin{center}
      \begin{tabular}{|c|c|c|c|c|}
                \hline & $a_{1}$ & $a_{2}$ & ... & $a_{r}$\\
                \hline $c_1$ & 1 & 0 &  & 1\\
                \hline $c_2$ & 1 & 1 &  & 1\\
                \hline ... & 0 & 0 &  & 1\\
                \hline $c_z$ & 1 & 1 & & 1\\
                \hline
                \end{tabular}\\
                \begin{flushleft}
                \begin{footnotesize}
                $a_0$ represents an attribute from dimension $d$
                and $r$ is the number of attributes in $C$.
                \end{footnotesize}
                \end{flushleft}
      \caption{Predicate term schematic table \label{table1}}
     \end{center}
\end{table}

\item \textbf{Candidate graph fragmentation}.
Each obtained predicate term and an additional predicate, called
ELSE, form an horizontal fragment. The ELSE predicate is the
negation of the conjonction of all predicate terms. It is added to
ensure fragmentation completeness. To ensure fragmentation
disjonction, a set of minterms is also created (Section
\ref{method1}).

\textbf{Example}. $t_1$ and ELSE=$\neg p_1$ or $\neg p_3$ or $\neg
p_5$ or $\neg p_2$ are predicate terms used to fragment
$G_{dimension_{customer}}$.

\end{enumerate}

\subsection{Fact fragmentation}
\label{fact}

The $G_{facts_f}$ graphs are finally fragmented according to
horizontal fragments obtained by applying either the PC or AB method
on dimensions. The fragmentation of $G_{facts_f}$ graphs is achieved
by semi-join operations based on a virtual key reference. This key
defines the relationships between $G_{dimension_{d}}$ and
$G_{fact_f}$ graphs. It is explicitly defined by the join
qualification expression provided in Figure~\ref{fig:exp} and
consists of a conjunction of two path expressions. These path
expressions check whether nodes in $G_{dimension_{d}}$ graphs
correspond to nodes in $G_{facts_f}$ graphs.

\begin{figure}[hbtp]
\centering
\footnotesize{\texttt{document($facts_f.xml$)/FactDoc/dimension[@dim-id=\\document($dimension_{d}.xml$)/dimension/Level/@id]}}\\
\footnotesize{\texttt{and}}\\
\footnotesize{\texttt{document($facts_f.xml$)/FactDoc/dimension[@value-id\\=document($dimension_{d
}.xml$)/dimension/Level[@id\\=@dim-id]/instance/@id]}} \caption{Join
qualification} \label{fig:exp}
\end{figure}

\begin{figure}[htbp]
{\centering
\resizebox*{0.3\textwidth}{!}{\includegraphics{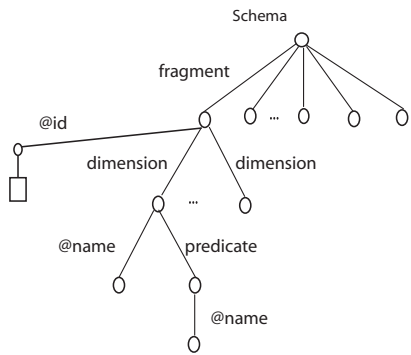}}
\par}
\caption{Fragmentation schema} \label{fig:schema}
\end{figure}

We finally build an XML document that represents the fragmentation
schema, \emph{fragmentation\_schema.xml}. Its corresponding graph,
denoted $Schema$, is provided in Figure~\ref{fig:schema}. The root
node, \emph{Schema}, is composed of \emph{fragment} elements
describing the obtained fragments. Each fragment is identified by an
\emph{@id} attribute and contains \emph{dimension} elements. A
dimension element is identified by a \emph{@name} attribute and
contains \emph{predicate}
elements that store minterms used for fragmentation.\\

\section{Experiments}
\label{sec:exps}

\subsection{Experimental conditions}

In order to validate our proposal experimentally, we use XWeB (the
XML Data Warehouse Benchmark) \cite{MahboubiD06}. XWeB is based on
the reference model defined in Section \ref{sec:xdwm}, and proposes
a test XML data warehouse and its associated XQuery decision-support
workload.

XWeB's warehouse consists of \emph{sale} facts characterized by the
amount (of purchased products) and quantity (of purchased products)
measures. These facts are stored in the $facts_{sales}.xml$ document
and are described by four dimensions: \emph{Customer},
\emph{Supplier}, \emph{Date} and \emph{Part} stored in the
$dimension_{Customer}.xml$, $dimension_{Supplier}.xml$,
\emph{dimension} $_{Date}.xml$ and \emph{dimension}$_{Part}.xml$
documents, respectively. XWeB's warehouse characteristics are
displayed in Table \ref{Table:dw-characteristics}.

\begin{table}[hbtp]
\begin{center}
\begin{tabular}{|l|l|}
\hline \small{\textbf{Facts}} & \small{\textbf{Number of cells}}\\
\hline Sale facts & 7000 \\
\hline
\hline \small{\textbf{Dimensions}} & \small{\textbf{Number of instances}}\\
\hline Customer & 1000 \\
\hline Supplier & 1000 \\
\hline Date  & 500 \\
\hline Part  & 1000 \\
\hline
\hline \small{\textbf{Documents}} & \small{\textbf{Size (MB)}}\\
\hline $facts_{sales}.xml$ & 2.14\\
\hline $dimension_{Customer}.xml$ & 0.431\\
\hline $dimension_{Supplier}.xml$ & 0.485\\
\hline $dimension_{Date}.xml$  & 0.104\\
\hline $dimension_{Part}.xml$  & 0.388\\
\hline
\end{tabular}
\caption{XWeB warehouse characteristics}
\label{Table:dw-characteristics}
\end{center}
\end{table}

XWeB's workload is composed of queries that exploit the warehouse
through join and selection operations. We extend this workload by
adding queries and selection predicates in order to obtain a
significant fragmentation. Our workload is available on-line
\footnote{\scriptsize{http://eric.univ-lyon2.fr/$\sim$hmahboubi/Workload/workload.xq}}.
We ran our tests on a Pentium 2 GHz PC with 1 GB of main memory and
an IDE hard drive under Windows XP. We use the X-Hive XML native
DBMS \footnote{\scriptsize{http://www.x-hive.com/products/db/}} to
store and query the warehouse.

\subsection{Experiments}

Our experiments measure workload execution time, with and without
using fragmentation and separately evaluate the PC and AB primary
fragmentation strategies (Section \ref{method1} and \ref{method2},
respectively). The fragments we achieve are stored in distinct
collections to simulate data distribution. Each collection can
indeed be considered as a distinct node/site and can be identified,
targeted and queried separately. To measure query execution time
over a fragmented warehouse, we first identify the required
fragments with the $Schema$ graph. Then, we execute the query over
each fragment and save execution time. To simulate a parallel
execution, we only consider the maximum execution time. We conducted
two series of experiments.

\subsubsection{First series of experiments}

This series of experiments helps observe the impact of data
warehouse size and workload characteristics on fragmentation
quality. For this purpose, we exploit three warehouse and workload
configurations (Table \ref{Table:configuration}) in which we vary
warehouse size (i.e., the number of facts) and the number of
workload queries and selection predicates.

\begin{table}[hbtp]
\begin{tabular}{|p{2.5cm}|c|c|c|}

\hline & \small{\textbf{Config. 1}} & \small{\textbf{Config. 2}} & \small{\textbf{Config. 3}}\\
\hline
%\multicolumn{4}{|c|}{\textbf{Data warehouse}}\\
%\hline
\hline \small{Number of facts} & 800 & 800 & 4000\\
\hline
%\multicolumn{4}{|c|}{\textbf{Workload}}\\
%\hline
\hline \small{Number of queries} & 13 & 19 & 19\\
\hline \small{Number of join operations} & 22 & 35 & 35\\
\hline \small{Number of predicates} & 20 & 30 & 30\\
\hline
\end{tabular}
\caption{Warehouse and workload configurations}
\label{Table:configuration}

\end{table}

%In all configuration, we obtain 159 fragments by using the PC and
%119 fragments with the AB strategy.

\begin{figure}[t]
{\centering
\resizebox*{0.4\textwidth}{!}{\includegraphics{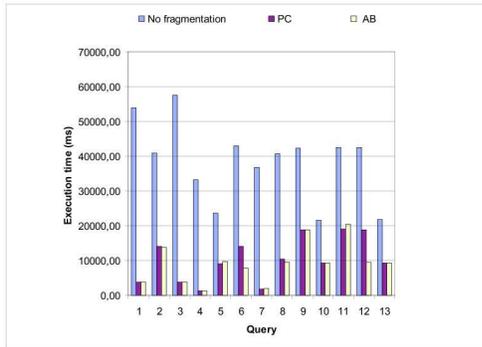}}
\par}
\caption{Configuration 1 results} \label{fig:exp1}
\end{figure}

\begin{figure}[h]
{\centering
\resizebox*{0.4\textwidth}{!}{\includegraphics{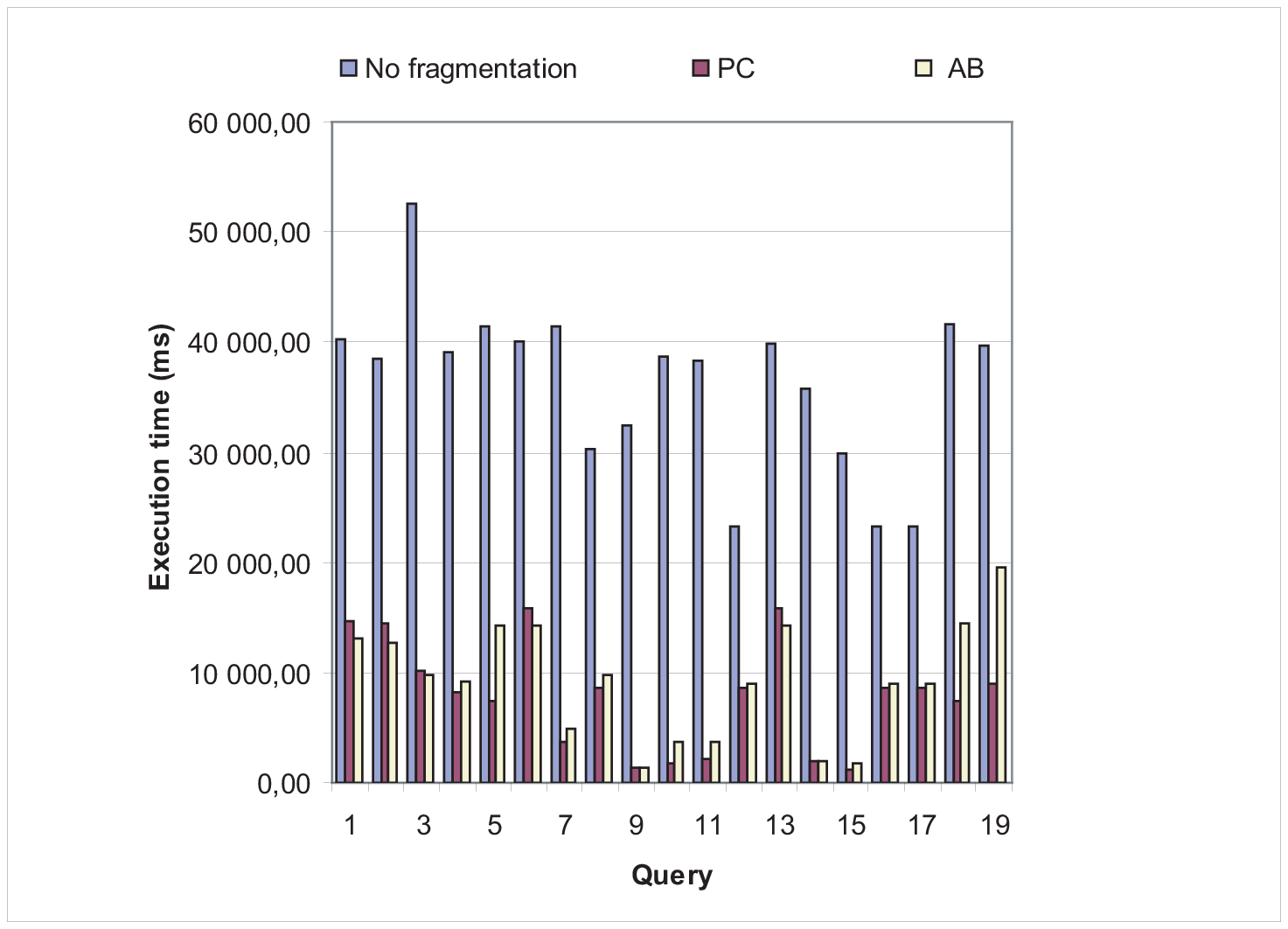}}
\par}
\caption{Configuration 2 results} \label{fig:exp2}
\end{figure}

\begin{figure}[t]
{\centering
\resizebox*{0.4\textwidth}{!}{\includegraphics{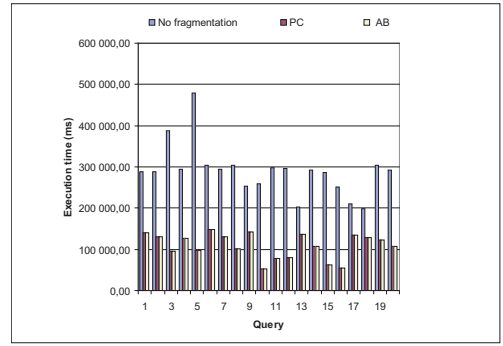}}
\par}
\caption{Configuration 3 results} \label{fig:exp3}
\end{figure}

Experiment results for configurations 1, 2 and three are showed on
Figures~\ref{fig:exp1}, \ref{fig:exp2} and \ref{fig:exp3},
respectively. In these figures, the X axis represents workload
queries and the Y axis features query execution time when no
fragmentation is applied on the warehouse, and when derived
horizontal fragmentation is applied with PC and AB primary
fragmentation.

For configuration 1, we obtain an average gain over no fragmentation
of 72.95\% with PC and 76.32\% with AB. For configuration 2, in
which the number of queries is increased over configuration 1, PC
improves query execution time by 74.53\% and AB by 78.32\% on
average. Finally, in configuration 3, we increase the number of
facts and obtain an average gain of 62.59\% with PC and 80,17\% with
AB.

These results confirm that fragmentation improves query performance.
They also show that AB provides more benefit than PC in all our test
cases. We think this is thanks to AB's use of query frequencies to
group in the same fragment all dimension instances and facts needed
to perform a given join operation. In addition, we notice that PC
fragmentation gain significantly declines in configuration 3, i.e.,
when warehouse size increases. We think this is due to the number of
fragments produced by PC, which is greater than that obtained with
AB (159 and 119, respectively). To further investigate this issue,
we conduct a second series of experiment where we further observe PC
and AB gain variation with respect to warehouse size.

\subsubsection{Second series of experiments}

\begin{figure}[h]
{\centering
\resizebox*{0.4\textwidth}{!}{\includegraphics{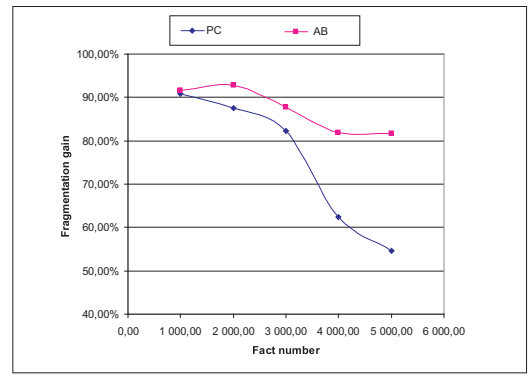}}
\par}
\caption{Fragmentation gain vs. warehouse size} \label{fig:exp5}
\end{figure}

This series of experiments helps aim at observing the effect of
warehouse size on fragmentation gain. We vary warehouse size from
1000 to 5000 facts and measure the fragmentation gain achieved when
using PC and AB primary fragmentation. The results of these
experiments are plotted in Figure~\ref{fig:exp5}, whose X axis
represents the number of facts and Y axis the corresponding gains
obtained by PC and AB primary fragmentation.

Experiment results show that fragmentation gain declines when
warehouse size decreases with both primary fragmentation methods.
This is expected, since fragments become bigger and bigger, inducing
a higher and higher scan cost when performing join operations.
However, we also observe that performance degradation is reasonably
slow for AB, while it is much steeper for PC. We believe that this
is because AB builds fragments containing data required to perform
the most frequent join operations in $W$, while storing less
frequently accessed data in the ELSE fragment. PC does not take this
aspect into account. It just groups in the same fragment data
accessed by one or more queries simultaneously. It also uses
minterms that may distribute data required to answer a single query
in different fragments, which multiplies reconstruction joins when
accessing data.

\section{Conclusion}
\label{sec:conclusion}

In this paper, we have adapt, to XML context, and compare the two
prevailing primary horizontal fragmentation methods from the
relational world, namely predicate construction and affinity-based
fragmentation. We have experimentally confirmed that derived
horizontal fragmentation helped improve query response time
significantly. Moreover, we also showed that affinity-based
fragmentation clearly outperformed predicate construction in all our
experiments, which had never been demonstrated before as far as we
know, even in the relational context.

The natural follow-up of this work is to distribute fragmented XML
warehouses on a data grid. This raises several issues that include
processing a global query into subqueries to be sent to the right
nodes in the grid, and reconstructing a global result from subquery
results. Properly indexing the distributed warehouse to guarantee
good performance shall also be very important.

\bibliographystyle{abbrv}
\bibliography{fragmentation}

\end{document}